\documentclass[aps,prl,twocolumn,
superscriptaddress,showpacs,amsmath,amssymb,unsortedaddress]{revtex4-1}
\usepackage{graphicx}% Include figure files
\usepackage{bm}% bold matha
\usepackage{epsfig}
\usepackage{color}
\begin{document}

\title{Laser Pulse Heating of Spherical Metal Particles}

\author{Michael I. Tribelsky}
\email[E-mail: ]{tribelsky_at_mirea.ru}
%\thanks{}
\affiliation{A.N.Nesmeyanov Institute of
Organoelement Compounds, Russian Academy of Sciences, Vavilova St. 28,  Moscow, 119991, Russia}
\affiliation{Moscow State Institute of Radioengineering, Electronics
and Automation (Technical University), 78 Vernadskiy Ave., Moscow
119454, Russia}
\affiliation{Max-Planck-Institut f\"ur Physik komplexer Systeme,
N\"othnitzer Str. 38, Dresden 01187, Germany}

\author{Andrey E. Miroshnichenko}
\affiliation{Nonlinear Physics Centre, Research School of
Physical Sciences and Engineering, Australian National University,
Canberra ACT 0200, Australia}

\author{Yuri S. Kivshar}
\affiliation{Nonlinear Physics Centre, Research School of
Physical Sciences and Engineering, Australian National University,
Canberra ACT 0200, Australia}

\author{Boris S. Luk'yanchuk}
\email[E-mail: ]{Boris_L_at_dsi.a-star.edu.sg}
\affiliation{Data Storage Institute, Agency for Science, Technology and
Research, Singapore 117608}

\author{Alexei R. Khokhlov}
\affiliation{A.N.Nesmeyanov Institute of
Organoelement Compounds, Russian Academy of Sciences, Vavilova St. 28,  Moscow, 119991, Russia}
\affiliation{M.V.Lomonosov Moscow State University, Faculty of Physics, Lenin Hills, 1, Bldg. 2,  Moscow, 119992, Russia}

\begin{abstract}
We consider a general problem of laser pulse heating of spherical metal particles with the sizes ranging from nanometers to millimeters. We employ the exact Mie solutions of the diffraction problem and solve heat-transfer equations to determine the maximum temperature at the particle surface as a function of optical and thermometric parameters of the problem. The main attention is paid to the case when the thermometric conductivity of the particle is much larger than that of the environment, as it is in the case of metal particles in fluids. We show that in this case at any given finite duration of the laser pulse the maximum temperature rise as a function of the particle size reaches an absolute maximum at a certain finite size of the particle, and we suggest simple approximate analytical expressions for this dependence which covers the entire range of variations of the problem parameters and agree well with direct numerical simulations.
\end{abstract}


%\date{\today}

\maketitle

\section{Introduction}

The problem of laser pulse heating of absorbing particles embedded in transparent liquid medium is important for different applications, including stimulation of chemical reactions, laser sintering, selective killing of pathogenical bacteria or cancer cells, etc. A metal nanoparticle excited by laser
pulse with the frequency close to its plasmon resonance efficiently converts electromagnetic energy into thermal energy, with dramatic raising of the temperature in the surrounding medium. As a result, this problem or its substantial parts were a subject of many recent papers in physics, biology, medicine and chemistry (see, e.g., Refs.~\cite{r01,r1,r2,r3,r4} to cite a few).

In spite of many experimental works on the subject, a theoretical understanding of this phenomenon is rather limited. In a standard approach, the problem is formulated as the study of heat transfer with a source (energy release in the particle) obtained as a solution of the corresponding diffraction problem. Such a problem does not have simple exact analytical solutions, and different studies employ either approximate analytical methods (valid for certain limiting cases only) or direct numerical calculations with the specified parameters. To the best of our knowledge, neither general solution of this problem applicable for a broad range of variations of the problem parameters, nor suitable analytical expressions are available in the literature.

On the other hand, for many applications it is highly desirable to obtain simple analytical expressions which describe the temperature at the surface of the particle $T_s$ in a broad range of the problem parameters, and for the particle sizes ranging from nanometers to millimeters. In the present paper
we solve this problem.  In particular, we obtain analytical expressions for the maximum temperature at the particle surface as a function of optical and thermometric parameters of the problem. Our results cover a wide range of the particle sizes from a few nanometers to millimeters, and they
are confirmed by direct numerical solutions.

The paper is organized as follows. In Sec.~II we formulate the problem and introduce {\em four major spatial scales}, namely the particle size, thickness of the skin layer, as well as the characteristic length of the heat diffusion in the particle and surrounding medium, which determine a variety of different cases discussed below. Our analytical results and estimates are summarized in Sec.~III for {\em twelve different cases} which include all possible combination of the four spatial scales. Section~IV summarizes our numerical results which are found to be in a good agreement with the analytical predictions. Finally, Sec.~V concludes the paper.

\section{Problem Formulation}

We consider linear heating of metal particles by a single laser pulse with duration $\tau$. We assume that the particle has the thermometric conductivity $\chi_p$  (typically $\chi_p \sim 0.1 - 1$~cm$^2$/s), and it is embedded into a fluid with the thermometric conductivity $\chi_f$  (typically $\chi_f \sim 10^{-3} - 10^{-2}$~cm$^2$/s), such that the condition $\chi_p \gg \chi_f$ always holds, and the heat transfer is limited entirely by the heat diffusion. We neglect convection processes owing to large characteristic time required for the convection to arise and develop.

Thus, the heating problem depends on the following set of parameters: the intensity of the laser pulse $I$, its duration $\tau$, frequency $\omega$, optical and thermometric constants of the particle and environment. As for the intensity $I$, owing to the linearity of the problem, the heating rate is just proportional to $I$, so the results obtained for a given $I$ may be easily recalculated for any other its value by a simple scale transformation, see below. As for the other parameters, if the particle is spherical with radius $R$ and the thickness of the skin layer $\delta$ (for metals in the optical region $\delta \sim 10^{-5}$~cm), we can define {\em just four characteristic spatial scales} of the problem, namely $R$, $\delta$, $\sqrt{\chi_f \tau}$, and $\sqrt{\chi_p \tau}$. According to the general principle of the dimensional analysis, interplay between these four scales determines the entire variety of heating regimes of the particle. All these regimes are discussed below one by one.

\section{Analytical results}

\subsection{Small particles}

For small particles, we assume that $R \ll \delta$, so that the heating occurs in the entire bulk of the particle rather than on its surface. Interplay between other scales allows to consider several different cases which we analyze below.

\textbf{Case S1:} $R \ll \delta \ll \sqrt{\chi_f \tau} \ll \sqrt{\chi_p \tau}$. In this case the incident light penetrates into the entire particle and the energy release should be proportional to the particle volume, i.e. the absorption cross-section of the particle $\sigma_{abs} \sim R^3$. Bearing in mind that the dimension of $\sigma_{abs}$ is cm$^2$, it is convenient to write
\begin{equation}\label{sigma}
    \sigma_{abs} = \alpha k R^3,
\end{equation}
where $\alpha$ is a dimensionless quantity, $k$ stands for the wavenumber of the incident light in the environmental fluid: $k = n_f \omega/c$. Here $n_f$ is purely real refractive index of the fluid, $\omega$ designates the frequency of the incident light and $c$ stands for the speed of light in vacuum.

In the simplest case of the Rayleigh scattering comparison of Eq.~\eqref{sigma} with the well-known expression for $\sigma_{abs}$ for a spherical small particle~\cite{LL} provides the following expression for $\alpha$:
\begin{equation}\label{alpha}
    \alpha  = \frac{{12\pi \varepsilon ''}}{{{{(\varepsilon ' + 2)}^2} + \varepsilon '{'^2}}},
\end{equation}
where $\varepsilon '$ and $\varepsilon ''$ stand for the real and imaginary parts of the complex relative dielectric permittivity of the particle ($\varepsilon = \varepsilon_p/n_f^2$).

%Let us first estimate

Next, we neglect spatial inhomogeneity of the heat sources, supposing that the volume density of the sources inside the particle is a constant equal to $\sigma_{abs}I(t)/V$, where $I(t)$ is the power density of the laser pulse (in W/cm$^2$) and V is the particle volume. The ground for this neglect is that the actual inhomogeneity in the heat sources is rather weak and it results even in weaker temperature inhomogeneities, owing to high rate of heat transfer in metals. It allows to replace the actual 3D heat transfer problem by its spherically-symmetric version.

Moreover, in what follows we are interested in \textit{estimates\/} of the maximal temperature of the particle surface, rather than in its exact calculations. For this reason we employ the spherically-symmetric problem formulation even for large particles, when the illuminated part of the particle obviously has temperature higher than that in shadow. %Validity of this approximation will be checked later on based upon comparison of the obtained analytical results with computer simulations of the full 3D problem.

Taking into account that inequality $R \ll \sqrt{\chi_f \tau}$ means the temperature field in the vicinity of the particle is quasi-steady (i.e., the term with the temporal derivative of $T$ in the heat conduction problem may be neglected relative to the terms with the spatial derivatives) we obtain that within the framework of the approximations made the temperature field is described by the following spherically-symmetric boundary-value problem:
\begin{eqnarray}
  & &\kappa_p\frac{1}{r^2}\frac{\partial}{\partial r}\left(r^2\frac{\partial T}{\partial r}\right) + \frac{3\sigma I(t)}{4\pi R^3} =0,\;\;\mbox{at}\;\; r<R \label{in} \\
  & & \frac{1}{r^2}\frac{\partial}{\partial r}\left(r^2\frac{\partial T}{\partial r}\right) = 0,\;\;\mbox{at}\;\; r>R \label{out} \\
  & & T(t,R-0) = T(t, R+0), \label{TR}\\
  & & \kappa_p \left(\frac{\partial T}{\partial r}\right)_{R-0}= \kappa_f \left( \frac{\partial T}{\partial r}\right)_{R+0},\label{grad_R} \\
  & & T\rightarrow 0 \;\;\mbox{at}\;\; r \rightarrow \infty, \label{granInfty}
\end{eqnarray}
where $\kappa_p$ and $\kappa_f$ are thermal conductivity of the particle and environment, respectively and $t$ in the dependence $T(r,t)$ plays a role of a parameter. Here and in what follows $T$ stands for the temperature rise from the room temperature, see Eq. \eqref{granInfty}.

Integration of Eqs. \eqref{in}--\eqref{granInfty} yields a parabolic temperature profile at $r<R$ and $T=T_s(t)R/r$ at $r>R$ with the surface temperature of the particle
\begin{equation}\label{Ts1}
    T_s = \frac{\sigma I(t)}{4\pi R\kappa_f} \equiv \frac{\alpha R^2kI(t)}{4\pi\kappa_f}.
\end{equation}
Note, the temporal dependence of $T_s$ coincides with that for $I(t)$, so that the maximal temperature is achieved at the maximum of the laser pulse.

\textbf{Case S2:} $R \ll \sqrt{\chi_f \tau} \ll \delta \ll  \sqrt{\chi_p \tau}$. The case corresponds to a quasi-steady field of the temperature both in the particle and in the fluid. From the point of view of heat conductivity, the case is identical to (S1), so that Eq.~\eqref{Ts1} is valid.

\textbf{Case S3:} $R \ll \sqrt{\chi_f \tau}   \ll  \sqrt{\chi_p \tau} \ll \delta$. The case is identical to (S1), and again Eq.~\eqref{Ts1} is valid.

\subsection{Large particles}

For large particle, we assume that $R \gg \delta$, so that the energy release occurs in a thin layer near the particle surface.

\textbf{Case L1} $\delta \ll R \ll \sqrt{\chi_f \tau} \ll \sqrt{\chi_p \tau}$.  In this case the absorption cross-section should be
proportional to the square of the linear size of the particle, i.e.,
\begin{equation}\label{sigma2}
    \sigma_{abs} = \pi R^2 Q_{abs},
\end{equation}
where $Q_{abs}$ stands for the dimensionless efficiency.

Regarding the temperature field, owing to the condition $R \ll \sqrt{\chi_f \tau} \ll \sqrt{\chi_p \tau}$ it is still quasi-steady. The difference between the previous case is that now a good approximation is the energy release at the surface of the particle. Then, the temperature field inside the particle should satisfy the homogeneous Laplace equation. The only non-singular solution of this equation is a constant profile, so that $T(r,t)$ at $r<R$ is reduced to $T(t)$, where time $t$ once again is regarded as a parameter. As for the temperature field in the environmental fluid, it keeps the same profile as that at $R \ll \delta$, which eventually [bearing in mind Eq.~\eqref{sigma2}] brings about the following expression for $T_s$:
\begin{equation}\label{Ts2}
     T_s = \frac{\sigma I(t)}{4\pi R \kappa_f} \equiv \frac{Q_{abs} RI(t)}{4\kappa_f}.
\end{equation}

\textbf{Case L2:} $ \delta \ll \sqrt{\chi_f \tau} \ll R \ll \sqrt{\chi_p \tau}$. In this case the field inside the particle is still quasi-steady, so that its $r$-dependence may be neglected, see above, case (L1). Regarding the field outside the particle, it is essentially $t$- and $r$-dependent. To determine $T_s$ we may employ the energy conservation law. For simplicity we consider a rectangular laser pulse with intensity $I_0$ and duration $\tau$. To a certain moment of time $t \leq \tau$ the energy $W$ absorbed by the particle is $\sigma_{abs}I_0 t.$

The absorbed energy is consumed to heat the particle to temperature $T_s$ and to heat an adjacent layer of the fluid. The former requires the energy $(4/3)\pi R^3 C_p\rho_p T_s$, the latter $4\pi R^2 2\sqrt{\chi_f t}C_f\rho_f T_s/2$, (to enhance accuracy of the estimate we have taken into account that the scale of a layer heated to time $t$ is $2\sqrt{\chi_f t}$~\cite{Carslow} and replaced the profile of the temperature in the heated layer by its mean value $T_s/2$). Here $C$ and $\rho$ stand for the specific heat and density of the particle (subscript $p$) and fluid (subscript $f$), respectively.

Equalizing $W$ to the consumed energy and considering the equality as an equation for unknown $T_s$, one easily derives
\begin{eqnarray}
 T_s(t) & = & \frac{{\sigma_{abs} I_0t}}{{\frac{4}{3} \pi R^3{C_p}{\rho _p} + 4\pi R^2\sqrt {{\chi _f}t} {C_f}{\rho _f}}} \nonumber \\
 & \equiv & \frac{{Q_{abs} I_0t}}{{\frac{4}{3} R{C_p}{\rho _p} + 4\ \sqrt {{\chi _f}t} {C_f}{\rho _f}}}.\label{Ts3}
\end{eqnarray}

The obtained $T_s(t)$ is a monotonic function of $t$, so the maximal temperature is achieved in the end of the laser pulse. Replacement $t \rightarrow \tau$ brings about the corresponding expression for the maximal temperature $T_{s(max)}$.

\textbf{Case L3:} $ \delta \ll \sqrt{\chi_f \tau} \ll \sqrt{\chi_p \tau} \ll R$. According to the employed problem formulation this case correspond to heating of infinite compound space whose left semi-space has the thermometric properties of the particle, the right one those of the fluid, and energy is released at the boundary between the semi-spaces. This problem is exactly solvable~\cite{Carslow}. The solution yields the following expression for the surface temperature:
\begin{eqnarray}
    {T_s(t)} & = &\frac{\sigma I_0}{2R^2\pi\sqrt \pi}\frac{\sqrt{\chi _p\chi _f t}} {\kappa _p\sqrt \chi _f  + \kappa _f\sqrt\chi _p}  \nonumber\\
    & \equiv & \frac{\alpha I_0}{2\pi\sqrt \pi}\frac{\sqrt{\chi _p\chi _f t}} {\kappa _p\sqrt \chi _f  + \kappa _f\sqrt\chi _p}\label{Ts4}
\end{eqnarray}
Once again $T_s$ occurs a monotonic function of time, so the maximal temperature is achieved in the end of the pulse, at $t = \tau$.

\textbf{Case L4:} $\sqrt{\chi_f \tau} \ll \delta \ll R \ll  \sqrt{\chi_p \tau}$. The surface absorption of light. The case is identical to (L2), so that Eq.~\eqref{Ts3} is valid.

\textbf{Case L5:}  $\sqrt{\chi_f \tau}  \ll \delta \ll  \sqrt{\chi_p \tau} \ll R$. The case is identical to (L3), and Eq.~\eqref{Ts4} is valid.

\subsection{Other cases}

\textbf{Case O1:} $\sqrt{\chi_f \tau} \ll R \ll \delta \ll  \sqrt{\chi_p \tau}$. From the viewpoint of the heat transfer the case is equivalent to (S2), but the absorption cross-section is described by Eq.~\eqref{sigma}. As a result
\begin{eqnarray}
    T_s(t) & = & \frac{{\sigma_{abs}I_0t}}{{\frac{4}{3}\pi R^3 {C_p}{\rho _p} + 4\pi R^2 {C_f}{\rho _f}}\sqrt{\chi _f t}} \nonumber \\
    & \equiv & \frac{{\alpha kI_0t}}{{\frac{4}{3}\pi {C_p}{\rho _p} + 4\pi{C_f}{\rho _f}}\frac{\sqrt{\chi _f t}}{R}}.\label{Ts6}
\end{eqnarray}

\textbf{Case O2:} $\sqrt{\chi_f \tau} \ll R  \ll  \sqrt{\chi_p \tau} \ll \delta$. The case is identical to (O1), and Eq.~\eqref{Ts6} is valid.

\textbf{Case O3:} $\sqrt{\chi_f \tau} \ll   \sqrt{\chi_p \tau} \ll R  \ll \delta$. According to the approximation used the energy is released inside the particle in a spatially-uniform manner. The case occurs analogous to(O1), and again Eq.~\eqref{Ts6} is valid.

\textbf{Case O4:} $\sqrt{\chi_f \tau} \ll   \sqrt{\chi_p \tau}   \ll \delta \ll R$. In this case the energy release occurs in a surface layer with the volume $4\pi R^2 \delta$. The condition $\sqrt{\chi_p \tau}   \ll \delta $ allows to neglect distortion of the temperature inside the layer by heat conductivity. Thus, the released energy is consumed to heat the mentioned layer and the layer of fluid with thickness $2\sqrt{\chi_f t}$. The energy balance yields the following expression for $T_s$:
\begin{eqnarray}
    T_s(t)& = & \frac{\sigma_{abs} I_0 t}{4\pi R^2(C_p\rho_p\delta + C_f\rho_f\sqrt{\chi_f t})} \nonumber \\
    & \equiv & \frac{Q_{abs} I_0 t}{4(C_p\rho_p\delta + C_f\rho_f\sqrt{\chi_f t})}.\label{Ts12}
\end{eqnarray}
\mbox{}\\
The maximal temperature is achieved in the end of the laser pulse, at $t = \tau$.

Thus the entire set of possible relation between the four length scales of the problem has been inspected.

\begin{figure}[tbp]
  \includegraphics[width=1.\columnwidth]{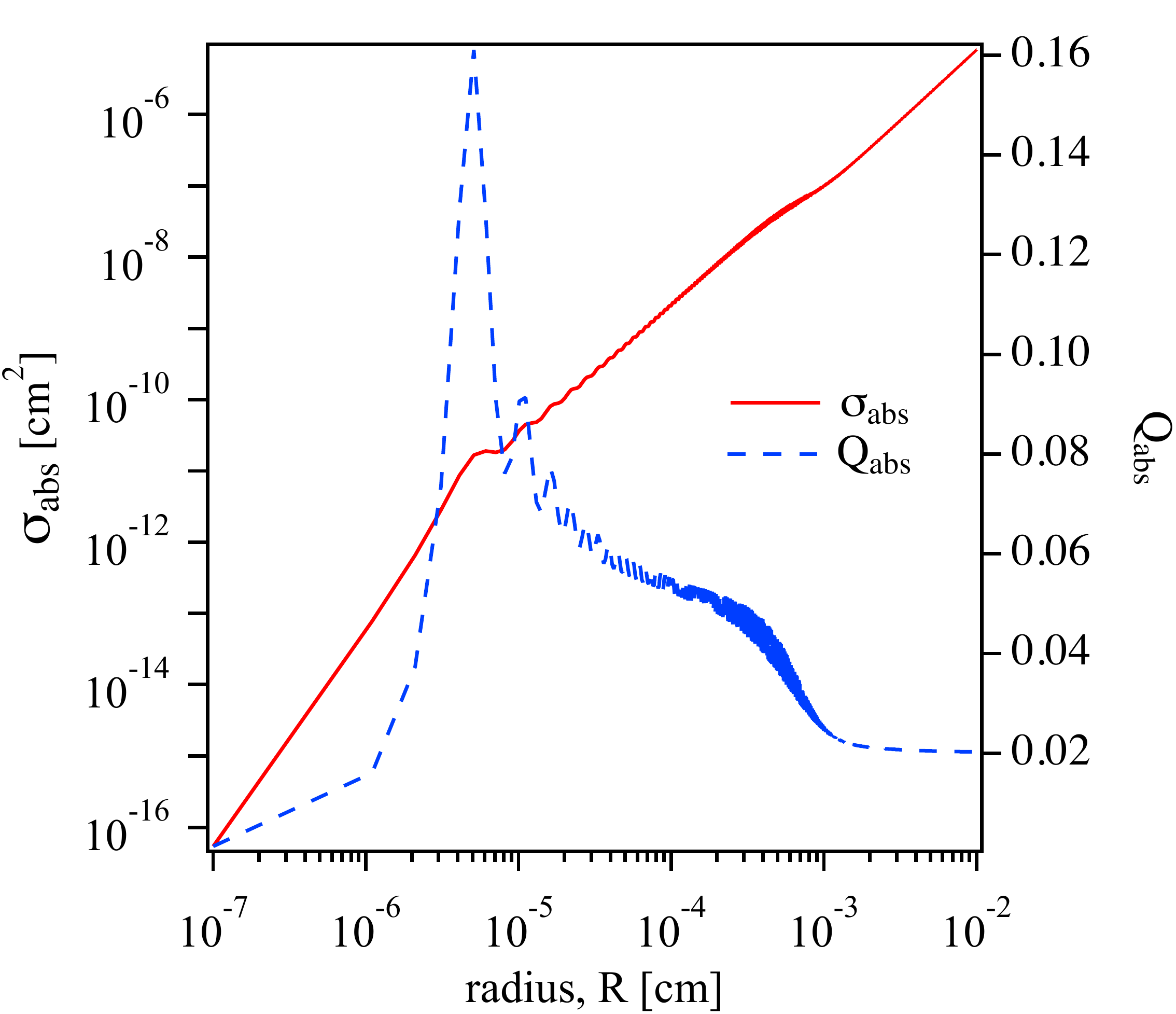}\\
  \caption{(color online) Absorption cross-section $\sigma_{abs}$ and the corresponding dimensionless efficiency $Q_{abs}$ for a spherical gold particle at $\lambda = 532$ nm and $\delta(\lambda)$ = 22 nm as functions of the particle radius $R$. Calculations according to the exact Mie solution.}\label{F1}
\end{figure}

\begin{figure}[tbp]
  \includegraphics[width=1.\columnwidth]{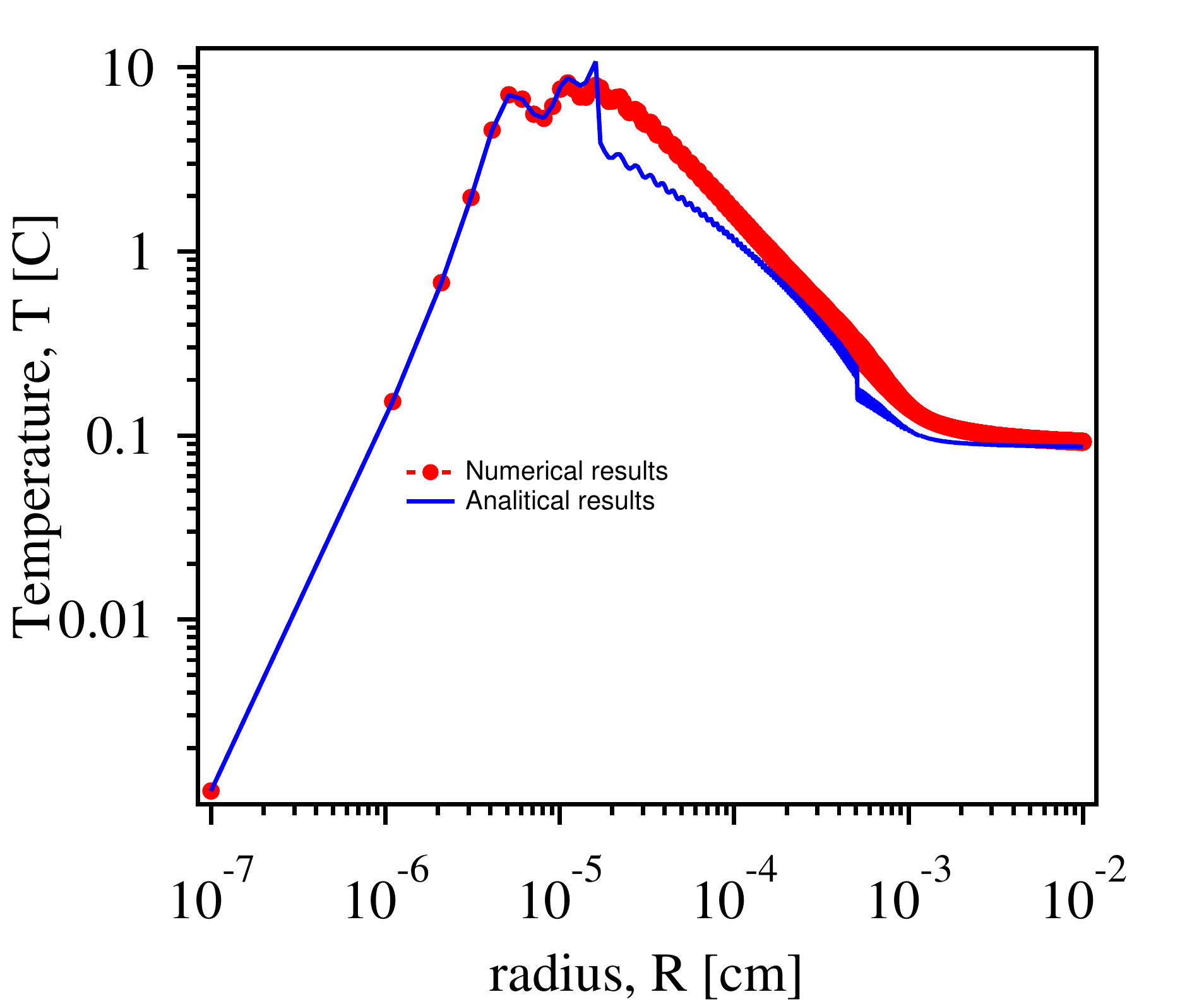}\\
  \caption{(color online) Comparison of the analytical and numerical results for the maximal temperature of a spherical gold particle with radius $R$ in water heated by a rectangular laser pulse with wavelength 532 nm, $I_0 = 2\cdot10^5$ W/cm$^2$ and $\tau = 50$ ns. [$\delta \ll \sqrt{\chi_f \tau}$, see cases (S1) and (L1)-(L3)]. }\label{F2}
\end{figure}

\begin{figure}[tbp]
  \includegraphics[width=1.\columnwidth]{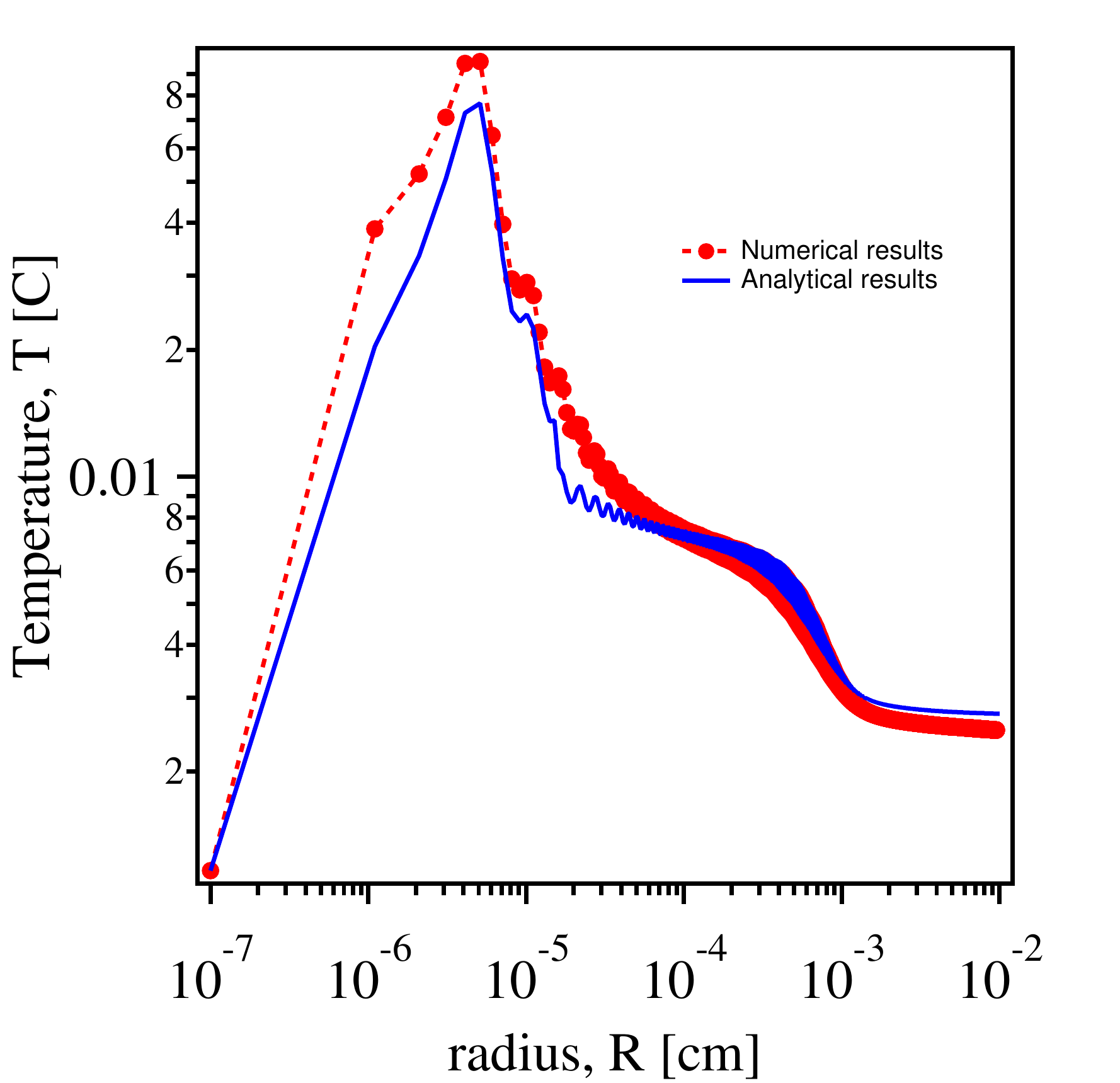}\\
  \caption{(color online) Same as in \protect{Fig. \ref{F2}} at $\tau = 50$ ps. [$\sqrt{\chi_f \tau} \ll \delta \ll \sqrt{\chi_p \tau}$, see cases (S2), (O1), and (L4)]. }\label{F3}
\end{figure}

\begin{figure}[tbp]
  \includegraphics[width=1.\columnwidth]{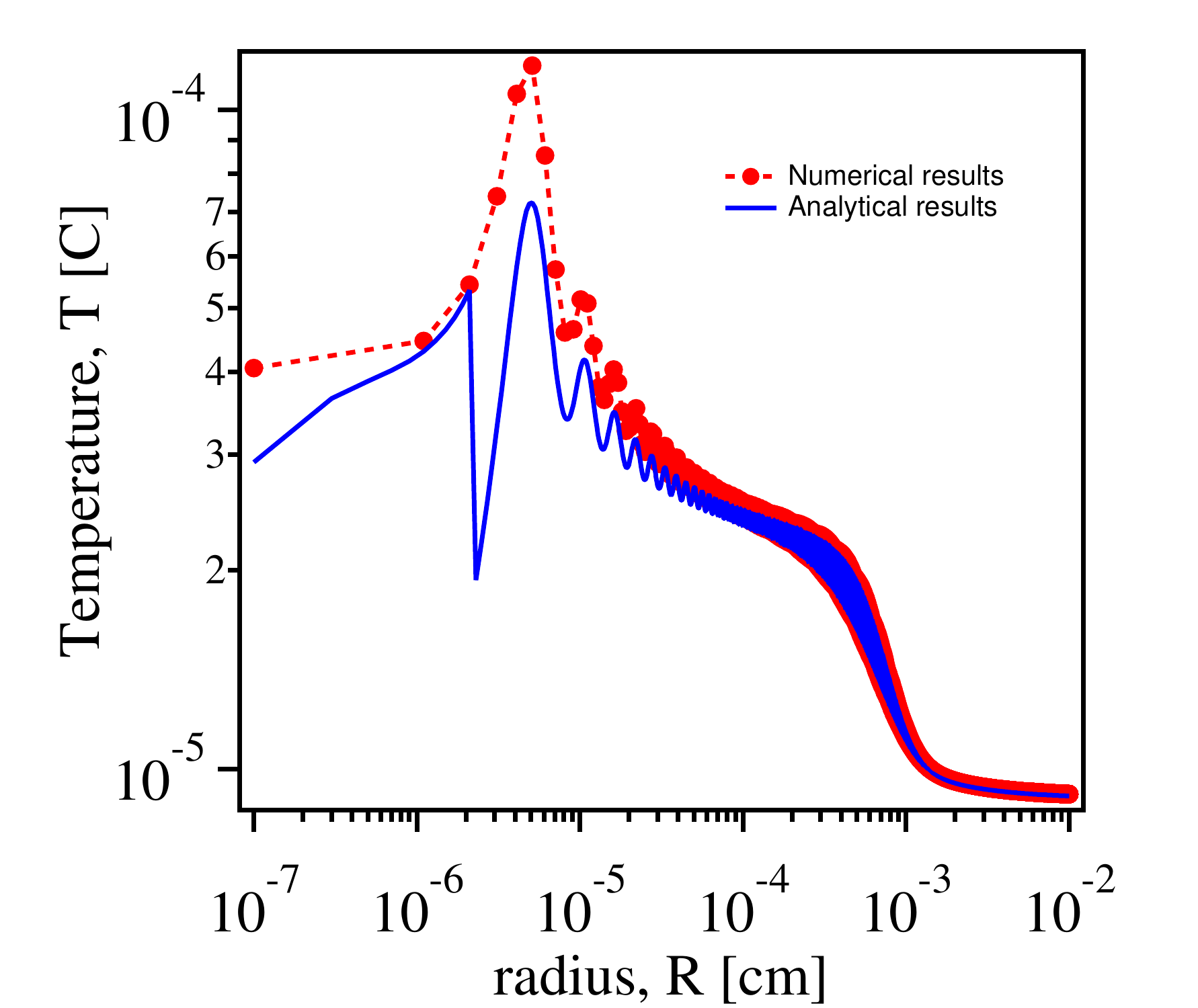}\\
  \caption{(color online) The same as that shown in \protect{Fig. \ref{F2}} at $\tau = 50$ fs. [$\sqrt{\chi_p \tau} \ll \delta$, see cases (S3), and (O2)-(O4)]. }\label{F4}
\end{figure}

\begin{figure}[tbp]
  \includegraphics[width=1.\columnwidth]{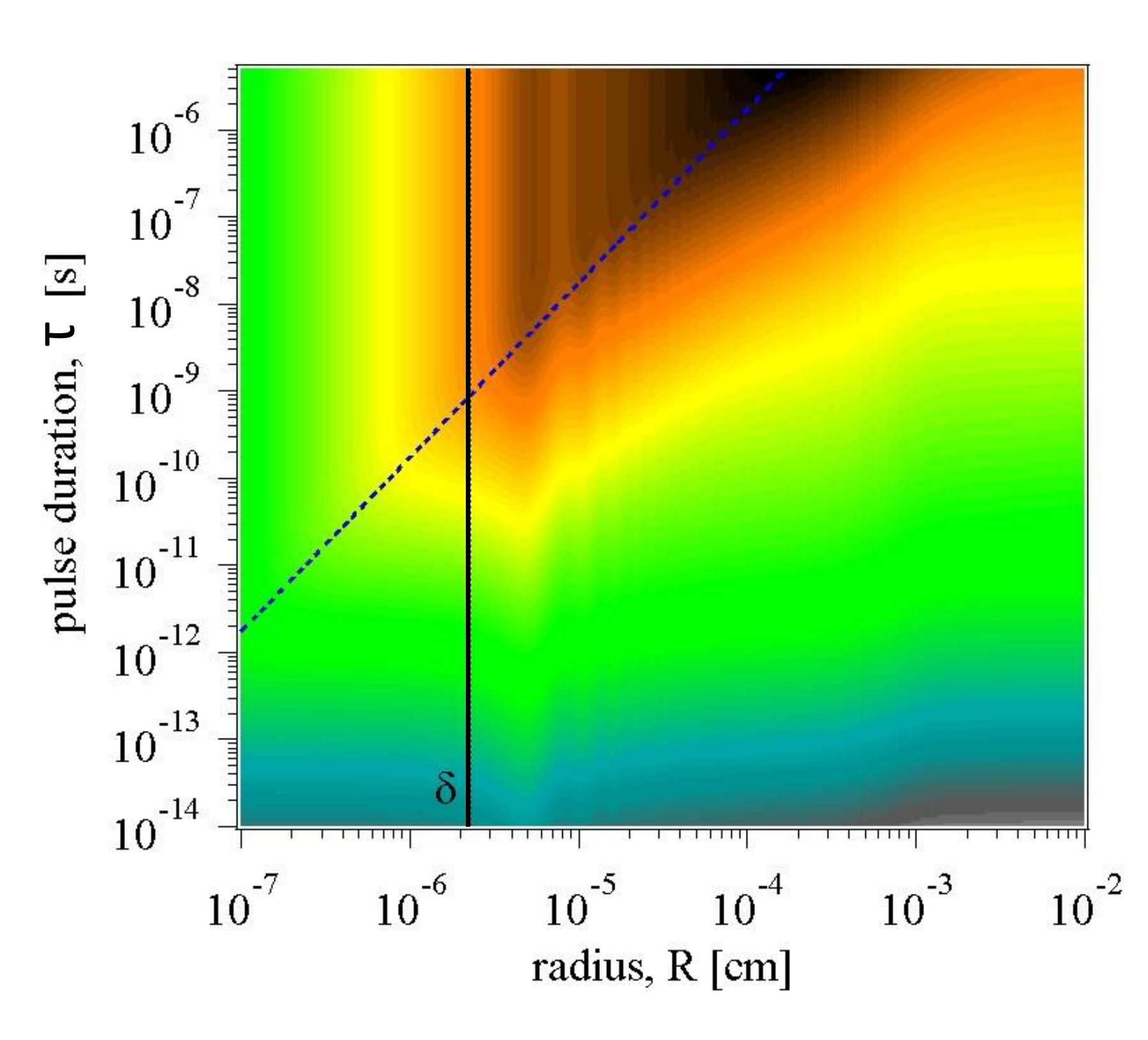}\\
  \caption{(color online) Density plot of the maximum of the surface temperature $T_{s(max)}$ (arbitrary units) for a gold particle in water irradiated by a rectangular laser pulse with \mbox{$I_0 = 2 \cdot 10^5$ W/cm$^2$} as a function of the duration of the pulse $\tau$ and the particle radius $R$. The dashed line corresponds to $R = 2\sqrt{\chi_f \tau}$. Above this line the temperature becomes $\tau$-independent, see Eqs.~\eqref{Ts1}, \eqref{Ts2}.}\label{F5}
\end{figure}

\begin{figure}[tbp]
  \includegraphics[width=1.\columnwidth]{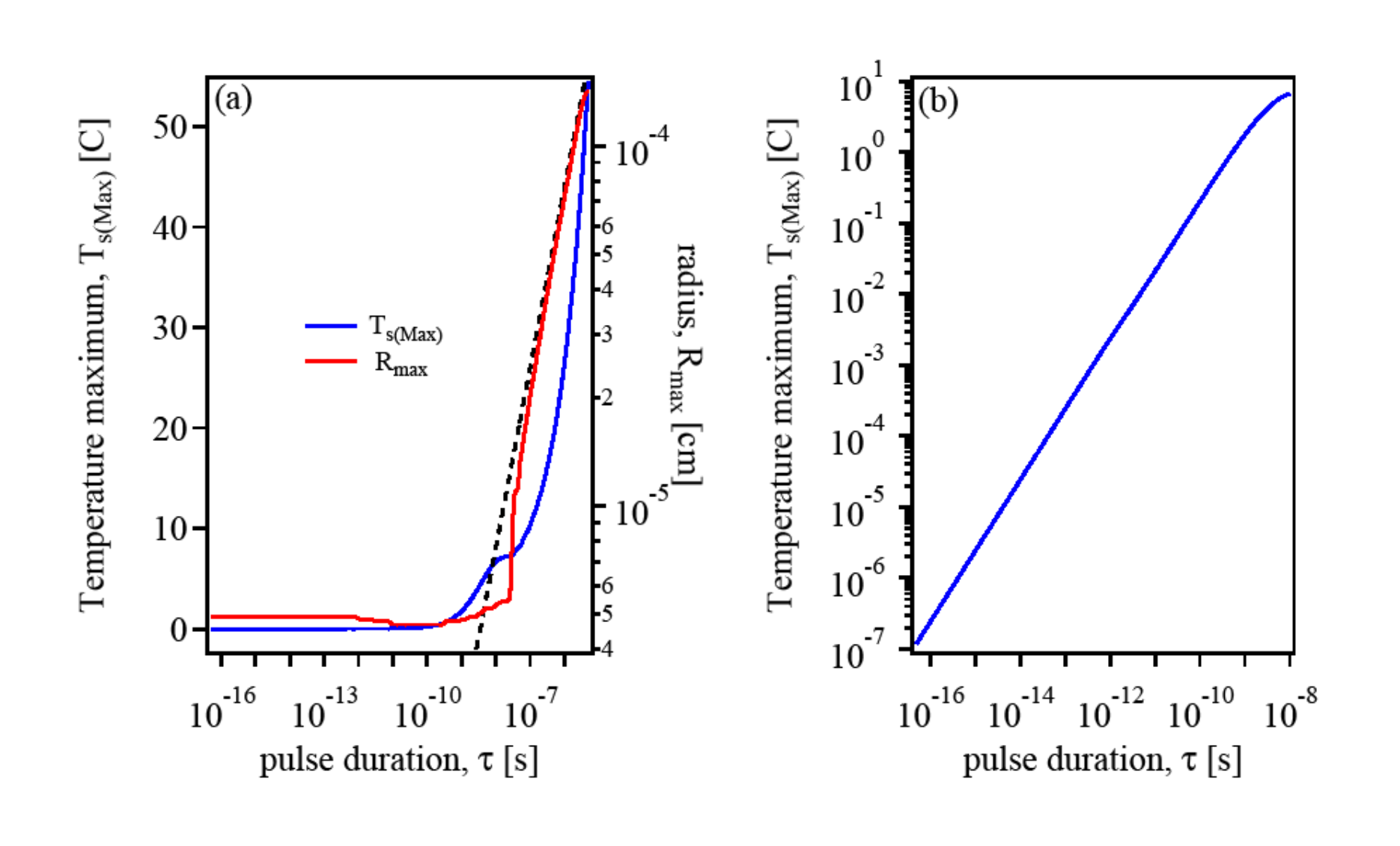}\\
  \caption{(color online) The radius $R_{max}$ providing the absolute maximum of the surface temperature and the corresponding temperature $T_{s(Max)}$ as functions of the laser pulse duration $\tau$ for a gold particle in water irradiated by a rectangular laser pulse with \mbox{$I_0 = 2 \cdot 10^5$ W/cm$^2$}. The dashed line indicates dependence $R = 2\sqrt{\chi_f \tau}$ (a). The initial part of plot $T_{s(Max)}(\tau)$, shown in panel (a), in Log-Log scale (b).}\label{F6}
\end{figure}

\section{Numerical vs. analytical results}

Let us discuss the results obtained. Note that the dependence of the dimensionless quantities $\alpha$ and $Q_{abs}$ on $R$ usually is very weak except for a narrow region centered about $R \sim \delta$, where the dependence $\sigma_{abs} \sim R^3$ is replaced by $\sigma_{abs} \sim R^2$. To illustrate this point the dependencies $\sigma_{abs}(R)$ and the corresponding $Q_{abs}(R)$ for a spherical gold particle are presented in Fig.~\ref{F1}. The calculations are made according to the exact Mie solution~\cite{Book} and actual optical constants of gold~\cite{Palik} at the wavelength of the incident light in a vacuum $\lambda = 532$ nm. The corresponding value of $\delta$ calculated as $c/(\omega n_{Au}'')$, where $n_{Au}''$ stands for the imaginary part of the refractive index of gold at the given wavelength, is 22 nm. The cubic dependence $\sigma_{abs}(R)$ [linear dependence of $Q_{abs}$, i.e. independence of $\alpha$ of $R$] at $R<\delta$ is seen straightforwardly. At $R>\delta$ the efficiency  $Q_{abs}$ drops from 0.6 to 0.2 when $R$ varies in three order of magnitude, i.e. the dependence $Q_{abs}(R)$ is extremely weak, and our assumption $\sigma_{abs} \sim R^2$ does capture the main $R$-dependence of $\sigma_{abs}$ in this area.

It means that the main dependence of $T_{s(max)}$ on the particle size is given by the explicit $R$-dependence of Eqs. \eqref{Ts1}, \eqref{Ts2}--\eqref{Ts12}. In particular, for cases (i)--(iv) the quadratic growth of $T_{s(max)}$ with an increase in $R$ is replaced by the linear at $R \sim \delta$, then $T_{s(max)}$ reaches its maximum at $R \sim \sqrt{\chi_f \tau}$, declines with further increase in $R$ and finally approaches to a constant at $R \gg \sqrt{\chi_p \tau}$. For other case $T_{s(max)}(R)$ is treated in a similar manner.

Accuracy of the developed approach is illustrated by comparison of the obtained analytical expressions with numerical simulations of the corresponding spherically-symmetric version of the heat conduction equation for a gold particle in water presented in Fig. 2--4, where $\sigma_{abs}$ is taken from the exact Mie solution for the gold particle. The particle is heated by a rectangular laser pulse, whose intensity for definiteness is assigned the typical value  $I_0 = 2\cdot10^5$ W/cm$^2$, and various values of $\tau$ (indicated in the figure captions).

Note, that there are two competing mechanisms of maximization of the surface temperature. The first is related to optics being associated with the local maximum of $Q_{abs}$ at $R \simeq \delta$, see Fig. \ref{F1}. The second is related to the heat transfer problem. It is associated with the change of the quasi-steady, $R$-dependent temperature field in the vicinity of the particle to the essentially time-dependent temperature profile, which does not depend on $R$. The change occurs at $R \approx 2\sqrt{\chi_f \tau}$.

To understand the relative role of these mechanisms we plot the $T_{s(max)}$ as a function of $R$ and $\tau$, see Fig.~\ref{F5}. At every given $R$ the $T_{s(max)}$ increases monotonically with an increase in $\tau$ until the latter reaches the values $\tau \simeq R^2/\chi_f$. Then, $T_{s(Max)}$ becomes $\tau$-independent.

Next, we perform the following calculations. For every given $\tau$ we find such a value of $R = R_{max}$ that maximizes $T_{s(max)}(R)$, i.e., provides the absolute maximum of the surface temperature, which may be achieved for the given $\tau$. Then, $R_{max}$ and the corresponding temperature $T_{s(Max)} = T_{s(max)}(R_{max})$ are plotted as functions of $\tau$, see Fig.~\ref{F6}. It is seen straightforwardly that for the problem in question heating for several degrees and high begins from $\tau \geq 10^{-9}$ s, when the heat transfer mechanism prevails over the optic one. Let us stress that, as it has been already pointed out in Sec. II, owing to the linearity of the problem the temperature is just proportional to $I_0$, which allows easily recalculate the results obtained for a given value of $I_0$ to any other its value.

To illustrate how our results may be employed in various applications let us consider an important example of selective laser photo-thermal therapy of cancer. Thus, it was shown that 40 nm gold nanoparticle conjugated to certain antibodies and then incubated with both human oral cancer cells and nonmalignant skin cells were preferentially and specifically bound to the cancer cells. Next, the
nanoparticle-labeled cells were exposed to a CW argon ion laser at 514 nm. It was found that the malignant cells required less than half the laser energy to be killed as compared to the benign cells. The destruction of the cancer cells occurred owing to laser heating of the gold nanoparticles up to a certain threshold temperature \cite{El-Sayed}.

On the other hand, applying this approach \textit{in vivo\/} to avoid unwanted effects one should minimize the exposure to the laser beam of the benign tissues and even the tumor itself, maximizing the rate of energy delivered to the nanoparticles and stored in the particles and their immediate vicinity. To this end a pulse laser should be employed. Application of our results indicates that reduction of the pulse duration from infinity (for a CW laser) to a nunosecond scale (without change of $I_0$) practically does not affect the maximal temperature rise of the nanoparticles. Presumably, heating of the nanoparticles by such a pulse still should be fatal for the cancer cells bound to them, while the exposure of the rest of the tumor and the benign tissues will be reduced dramatically.

%In all the cases $T_{s(max)}$ exhibits a pronounced maximum at $R \approx 2\sqrt{\chi_f \tau}$. The reason for that is as follows. At $R \ll \sqrt{\chi_f \tau}$ the temperature gradient in the environmental fluid in the vicinity of the particle surface is proportional to $1/R$. At $R \gg \sqrt{\chi_f \tau}$  the gradient becomes $R$-independent, and the heated layer of the fluid has the width $\sqrt{\chi_f\tau}$. It affects the $R$-dependence of the net heat transfer from the heated particle to the fluid. For more details see the above discussion of cases (i)-(xii). A crossover from one asymptotic to the other occurs just at $R \sim \sqrt{\chi_f \tau}$.
%
%Finally, the accuracy of the employed spherically-symmetric approximation has been checked against comparison of this version of the heat conductivity problem with numerical solutions of the actual 3D problem, where the energy release in the particle was calculated based upon the Maxwell equations. The comparison at several characteristic values of the problem parameters indicates that the spherically-symmetric approximation always provides for $T_{s(max)}$ values, which keep the same order of magnitude as those obtained from the solutions of the full 3D problem.

\section{Conclusion}

We have demonstrated that rather a complex problem of laser pulse heating of spherical metal particles embedded in a transparent fluid
may be described by relatively simple analytical expressions, which provide the dependence of the maximum temperature at the particle
surface $T_{s(max)}$ on the particle size and other parameters of the problem. We have demonstrated, by a direct comparison with the numerical simulations, that these expressions remain valid for any practically important size of the particle and duration of the laser pulse. More importantly,
at fixed values of the material constants, the function $T_{s(max)}(R)$ reaches local maxima at $R \simeq \delta$ and $R \sim \sqrt{\chi_f \tau}$. For a gold particle in water considerable heating begins from $\tau \geq 10^{-9}$ s, when the absolute maximum of the temperature is achieved at $R \approx 2\sqrt{\chi_f \tau}$. We believe these results not only give the most general solution of the problem but will also be useful to optimize the efficiency of laser pulse heating of nanoparticles in various problems of laser-matter interaction.

\section*{Acknowledgements}

The work was partially supported by the Australian Research Council. M.T. thanks the Max-Planck-Institut f\"ur Physik komplexer Systeme for kind hospitality during the substantial part of this project.

\end{document}